\documentclass[twoside,12pt]{article}
\usepackage{indentfirst}
\usepackage{bm}
\usepackage{epsfig}
\usepackage{graphicx}
\usepackage{color}
\usepackage{amsmath}
\usepackage{amssymb}
\usepackage{hyperref}

\newcommand{\abs}[1]{\ensuremath{\left| #1 \right|}}
\newcommand{\ket}[1]{{\vert #1\rangle}}

\newcommand{\braket}[2]{\langle#1\vert#2\rangle}

\newcommand{\ud}{\mathrm{d}}

\begin{document}


\thispagestyle{empty} \vspace*{0.8cm}\hbox
to\textwidth{\vbox{\hfill\huge\sf Commun. Theor. Phys.\hfill}}
\par\noindent\rule[3mm]{\textwidth}{0.2pt}\hspace*{-\textwidth}\noindent
\rule[2.5mm]{\textwidth}{0.2pt}


\begin{center}
\LARGE\bf General interaction quenches in a Luttinger liquid
\end{center}

\footnotetext{\hspace*{-.45cm}\footnotesize $^\dag$ E-mail: luht@lzu.edu.cn }

\begin{center}
\rm You-Ming Wei$^{1,2}$ \ and \ Hantao Lu$^{1,2\dagger}$ 
\end{center}

\begin{center}
\begin{footnotesize} \sl
${}^{1}$ School of Physical Science and Technology $\&$ Key Laboratory for Magnetism and Magnetic Materials of the MoE, Lanzhou University, Lanzhou 730000, China \\
${}^{2}$ Lanzhou Center for Theoretical Physics, Key Laboratory of Theoretical Physics of Gansu Province, Lanzhou University, Lanzhou, Gansu 730000, China
\end{footnotesize}
\end{center}


\vspace*{2mm}

\begin{center}
\begin{minipage}{15.5cm}
\parindent 20pt\footnotesize
We discuss a general interaction quench in a Luttinger liquid described by a paired bosonic Hamiltonian. By employing $\mathsf{su}(1,1)$ Lie algebra, the post-quench time-evolved wavefunctions are obtained analytically, from which the time evolution of the entanglement in momentum space can be investigated. We note that depending on the choice of Bogoliubov quasiparticles, the expressions of wavefunctions, which describe time-evolved paired states, can take different forms. The correspondence between the largest entanglement eigenvalue in momentum space and the wavefunction overlap in quench dynamics is discussed, which generalizes the results of D\'ora {\em et al} [2016, {\em Phys. Rev. Lett.} \textbf{117}, 010603]. A numerical demonstration on an XXZ lattice model is presented via the exact diagonalization method. 
\end{minipage}
\end{center}

\begin{center}
\begin{minipage}{15.5cm}
\begin{minipage}[t]{2.3cm}{\bf Keywords:}\end{minipage}
\begin{minipage}[t]{13.1cm}
quantum quench, exact diagonalization, Luttinger liquid, XXZ model
\end{minipage}\par\vglue8pt

\end{minipage}
\end{center}

\section{Introduction}\label{sec:intro}

Quantum quench, where parameters in the Hamiltonian are changed discontinuously to leave system out of equilibrium, is an important subject in the study of nonequilibrium dynamics of isolated quantum systems~\cite{Polkovnikov:2011, Mitra:2017}. Though the problem is well defined, most of the known exact results essentially come from noninteracting systems, e.g., the transverse field Ising model~\cite{suzuki2012quantum, Vidmar:2016}, and some pairing models~\cite{Yuzbashyan:2005a, Yuzbashyan:2005b, Yuzbashyan:2006dc}. For many-body systems the situation is much more complicated, and generally completely exact treatments are far beyond reach even for relatively simple models~\cite{Essler:2016, Caux:2016, Ilievski:2016}. In order to procure some qualitative understandings and to make decent quantitative predictions, we usually have to resort to effective theories. 

Analogous to the prevalent Landau's Fermi liquid theory, a bunch of one-dimensional (1D) quantum systems can be classified into a universal category known as {\em Luttinger liquids}~\cite{haldane1981luttinger, Giamarchi:2004, Gogolin:2004vy}, where only a handful of parameters are sufficient to address the low-temperature properties of systems effectively. A bit of surprise is that within the framework of the Luttinger liquid theory, it is possible to produce highly accurate descriptions on the nonequilibrium dynamics of the systems in their gapless (critical) phases. This observation has been justified by numerical computations on various lattice models~\cite{Karrasch:2012, Collura:2015}. 

One advantage of the investigations from the effective theory side is that analytical approaches, in many cases, can be carried out at least approximately~\cite{Cazalilla:2006, Cazalilla:2016}. In the work of D\'ora {\em et al}~\cite{Dora:2016}, based on the analysis of an effective Luttinger Hamiltonian, an interesting observation was made, which unveiled the correspondence between the time-evolved entanglement spectrum and the nonequilibrium characteristics in the quench dynamics, including the identification of the largest momentum-space entanglement eigenvalue and the Loschmidt echo. In their work, the correspondence was made plain first via the calculations on the simple Luttinger liquid Hamiltonian, and later was ratified by the numerical results on spin-chain lattice models. In our opinion, the work highlights the effectiveness of the Luttinger liquid theory in the nonequilibrium dynamics of 1D critical systems.

Nevertheless, in the paper~\cite{Dora:2016} the authors discussed primarily the noninteracting-to-interacting quenches, leaving the issue of general quenches untouched, e.g., a quench from one nonzero coupling strength to another. In this study, we would like to address the quench dynamics of the Luttinger liquid model in a bit of more general settings, by investigating the quenches that can start from an arbitrary (interacting) ground state (GS). With the assistance of $\mathsf{su}(1,1)$ algebra, we procure the exact expression of the time-evolved wavefunction, and obtain the largest entanglement eigenvalue in momentum space accordingly. We note that the correspondence raised in \cite{Dora:2016} can be generalized consequently. We confirm it numerically in the XXZ-related model [see (\ref{eq:XXZ3}) for the Hamiltonian], which is a lattice realization of the Luttinger Hamiltonian~(\ref{eq:H1}). There we find that the agreement holds in the sense that the largest momentum-space entanglement eigenvalue coincides with the overlap of the time-evolved wavefunction with its noninteracting GS, rather than with the initial interacting GS from which the quench starts. 

The rest of the paper is organized as follows. After an introduction of the Luttinger liquid Hamiltonian in terms of paired bosons in section~\ref{sec:pairedH}, we discuss the general $g_0\to g$ quench and calculate the aftermath time-evolved wavefunction in section~\ref{sec:g0g}. We note specifically that the expression of the time-evolved wavefunction depends on the choice of Bogoliubov quasiparticles. The correspondence between the wavefunction overlaps, including the Loschmidt echo, and the largest entanglement eigenvalues, is examined in section~\ref{sec:LE}. Numerical results on the fermionic lattice model are presented in section~\ref{sec:numerics}. The discussion and conclusion is given in section~\ref{sec:conclusion}.

\section{The paired bosonic Hamiltonian}\label{sec:pairedH}

As a description of a 1D Luttinger liquid, the Hamiltonian we consider here takes the following form~\cite{Dora:2016}:
\begin{equation}
H=\sum_{q\neq 0}\omega(q)b^{\dagger}_qb_q+\frac{g(q)}{2}\left[b_q^{\dagger}b_{-q}^{\dagger}+b_qb_{-q}\right],
\label{eq:H1}
\end{equation}
where $b_q\ (b_q^{\dagger})$ is the annihilation (creation) operator of a bosonic particle with momentum $q$. For Luttinger liquids, the dispersion relation and the pairing strength are given as~\cite{Giamarchi:2004} 
\begin{equation}
\omega(q)=v\abs{q},\qquad g(q)=g\abs{q}.
\label{eq:wg}
\end{equation}

The above Hamiltonian can be diagonalized via the Bogoliubov transformation: 
\begin{equation}
(a_q\quad a_{-q}^{\dagger})=(b_q \quad b_{-q}^{\dagger})
\left(\begin{array}{cc} u(q) & v(q) \\ v^*(q) & u^*(q)\end{array}\right),
\label{eq:uv}
\end{equation}
where the condition of $\abs{u}^2-\abs{v}^2=1$ is imposed to preserve the bosonic commutations for the newly introduced $a$-operators. Since $\omega(q)$ and $g(q)$ are real functions, $u$, $v$ can be set to be real and parameterized by $\theta(q)$ as 
\begin{equation}
u(\theta)=\cosh(\theta/2),\qquad v(\theta)=\sinh(\theta/2), \qquad \tanh\theta(q)=g(q)/\omega(q).
\label{eq:uvtheta}
\end{equation}
Under the above Bogoliubov transformation~(\ref{eq:uvtheta}) (the condition of $\abs{g}<\abs{\omega}$ need to be met in order to guarantee the existence of solution), the pairing Hamiltonian~(\ref{eq:H1}) can be transformed into a diagonal form in terms of the Bogoliubov operators $a_q$:
\begin{equation}
H=\sum_{q>0}\epsilon(q)\left[a^{\dagger}_{q}a_{q}+a_{-q}a^{\dagger}_{-q}\right],
\label{eq:H3}
\end{equation}
where $\epsilon(q)=\sqrt{\omega^2(q)-g^2(q)}$. We note that for $\omega(q)$ and $g(q)$ given by (\ref{eq:wg}), $\theta$ is actually $q$ independent. For simplicity, in the new Hamiltonian (\ref{eq:H3}) the constant term $-\sum_{q>0}\omega_q$ has been dropped compared to the original Hamiltonian~(\ref{eq:H1}).

The Bogoliubov Hamiltonian~(\ref{eq:H3}) indicates that the ground state (GS) of the system should be the vacuum state of $a$-operators, i.e., $a_{q}\ket{\mathrm{GS}}=0$ is satisfied for each $q$. It is easy to verified that $\ket{\mathrm{GS}}$ can be expressed as a pairing state of the constituent particles as:
\begin{equation}
\ket{\mathrm{GS}}=\prod_{q>0}\frac{1}{\abs{u(q)}}e^{-\frac{v(q)}{u(q)}b_q^{\dagger}b_{-q}^{\dagger}}\ket{0},
\label{eq:GS}
\end{equation}
where $1/\abs{u(q)}$ serves as the normalized factor. Throughout this paper, $\ket{0}$ is reserved to specify the vacuum state of the original $b$-operator, i.e., $b_q\ket{0}=0$. 

\section{The $g_0\to g$ quench}\label{sec:g0g}

The issue we are going to address is the quench dynamics on the pairing strength $g$: starting from an interacting (entanglement) GS with respect to some (nonzero) value $g_0$, we switch the interaction strength abruptly from $g_0$ to another $g$ value. What is the post-quench dynamics of the system? 

For the Hamiltonian considered here, as have been mentioned before, the explicit form of the time-evolved wavefunction for the quench from the noninteracting limit (i.e., $g=0$) to a finite $g$ has been obtained, together with a discussion of the correspondence between the Loschmidt echo and the entanglement in momentum space~\cite{Dora:2013, Dora:2016}. Here in this section by utilizing the $\mathsf{su}(1,1)$ Lie algebra, we make a further step by calculating the post-quench wavefunction for the general case of $\forall \ g_0\to g$ (under the condition of $\abs{g_0}$ and $\abs{g}<\abs{\omega}$). 

The GS for the pairing strength $g_0$, denoted as $\ket{\mathrm{GS}_0}$, is already given by (\ref{eq:GS}). If we set the quench moment $t=0$, then
\begin{equation}
\ket{\Psi(0)}=\ket{\mathrm{GS}_0}=\sqrt{1-\lambda_0^2}e^{-\lambda_0b^{\dagger}\underline{b}^{\dagger}}\ket{0},
\label{eq:GS2}
\end{equation}
where $\lambda_0:=v_0/u_0$, and the Bogoliubov coefficients $u_0$, $v_0$ are determined by (\ref{eq:uvtheta}) for the given $g_0$. Since $(q,-q)$-pairs with different $q\ (q>0)$ in the Hamiltonian~(\ref{eq:H1}) are indeed decoupled from each other, for notation simplicity we would like to drop the label of $q$ in the formalism as long as there is no misunderstanding. The summation or multiplication over $q\,(>0)$ can be restored easily if necessary. For instance, $b^{\dagger}\underline{b}^{\dagger}$ usually stands for $\sum_{q>0}b_q^{\dagger}b_{-q}^{\dagger}$, as in the case of Eq.~(\ref{eq:GS2}).

For the quench $g_0\to g$ at $t=0$, the later-time wavefunction of the system is given by $\ket{\Psi(t)}=U(t)\ket{\Psi(0)}$, with the time-evolution operator $U(t)=e^{-iH(g)t}$ ($\hbar=1$). Note that $H(g)$ can be expressed in a diagonal form by its corresponding Bogoliubov quasiparticles [see (\ref{eq:H3})], and consequently, $U(t)=\exp[-i\epsilon t(a^{\dagger}a+\underline{a}\underline{a}^{\dagger})]$. In this situation we have two sets of Bogoliubov quasiparticle operators, i.e., $a_0$ and $a$, corresponding to $H(g_0)$ and $H(g)$, respectively. It can be shown that they are related by a composition of two successive Bogoliubov transformations [see (\ref{eq:uv})]:
\begin{equation}
(a\quad \underline{a}^{\dagger})=(a_0 \quad \underline{a}_{0}^{\dagger})
\left(\begin{array}{cc} \tilde{u} & \tilde{v} \\ \tilde{v} & \tilde{u}\end{array}\right),
\label{eq:uvtilde}
\end{equation}
where $\tilde{u}=\cosh\tilde{\theta}/2$, $\tilde{v}=\sinh\tilde{\theta}/2$, and $\tilde{\theta}=\theta-\theta_0$, with $\theta$ and $\theta_0$ the parameters for the Bogoliubov transformations~(\ref{eq:uvtheta}) for $H(g)$ and $H(g_0)$, respectively.

Equipped with the relation of $a\leftrightarrow a_0$~(\ref{eq:uvtilde}) in hand, we can write down an alternative form of $U(t)$ in terms of $a_0$ operators:
\begin{equation}
U(t)=e^{xK_{+}+yK_{-}+zK_{0}},
\label{eq:Ut}
\end{equation}
where we have introduced $K$-operators as
\begin{equation}
K_0=\frac{1}{2}\left[a_0^{\dagger}a_0+\underline{a}_0\underline{a}_0^{\dagger}\right],
\qquad K_{+}=a_0^{\dagger}\underline{a}_0^{\dagger},\qquad K_{-}=a_0\underline{a}_0,
\label{eq:Kvsa}
\end{equation}
and the coefficients in the exponents read
\begin{subequations}
\begin{equation}
x=y=-2i\epsilon t\tilde{u}\tilde{v}=-i\epsilon t\sinh\tilde{\theta}, \\
\end{equation}
\begin{equation}
z=-2i\epsilon t\left(\tilde{u}^2+\tilde{v}^2\right)=-2i\epsilon t\cosh\tilde{\theta}.
\end{equation}
\label{eq:xyz}
\end{subequations}
Note that the three $K$s satisfy the $\mathsf{su}(1,1)$ algebra~\cite{Truax:1985}:
\begin{equation}
[K_0,\ K_{\pm}]=\pm K_{\pm},\qquad [K_{+},\ K_{-}]=-2K_0.
\label{eq:su11}
\end{equation}
By employing the results (\ref{eq:Cs}) in~\ref{appA}, the time-evolution operator $U(t)$ in (\ref{eq:Ut}) can be factorized into a product form as
\begin{equation}
U(t)=e^{C_{+}(t)K_{+}}e^{C_{0}(t)K_{0}}e^{C_{-}(t)K_{-}},
\label{eq:Ut2}
\end{equation}
where particularly,
\begin{equation}
C_{+}(t)=\frac{-i\sin(\epsilon t)\sinh\tilde{\theta}}{\cos(\epsilon t)+i\sin(\epsilon t)\cosh\tilde{\theta}}.
\label{eq:C+}
\end{equation}
Then the normalized post-quench time-evolved wavefunction in terms of the $a_0$-Bogoliubov quasiparticles can be written as
\begin{equation}
\ket{\Psi(t)}=U(t)\ket{\Psi(0)}\sim \sqrt{1-\abs{C_{+}}^2}e^{C_{+}a_0^{\dagger}\underline{a}_0^{\dagger}}\ket{\mathrm{GS}_0}.
\label{eq:WF1}
\end{equation}
Here for simplicity we have dropped the phase factor owing to the action of $e^{C_{0}K_{0}}$ on $\ket{\mathrm{GS}_0}$, since it is not essential in our discussions. 

On the other hand, if we are instead interested in the expression of $\ket{\Psi(t)}$ in terms of the original creation (annihilation) operator $b^{\dagger}$ ($b$) with respect to its vacuum $\ket{0}$, we can employ the following procedure. First note that according to (\ref{eq:GS3}), the GS of $H(g_0)$ can be also expressed by an $a_0$-pair generating operator acting on the vacuum $\ket{0}$:
\begin{equation}
\ket{\mathrm{GS}_0}=\frac{1}{\sqrt{1-\lambda_0^2}}e^{-\lambda_0K_{+}}\ket{0},
\end{equation}
where $K_{+}$ is defined in (\ref{eq:Kvsa}). Combined with the result in (\ref{eq:WF1}), we have
\begin{equation}
\ket{\Psi(t)}\sim e^{C_{+}K_{+}}e^{-\lambda_0K_{+}}\ket{0}=e^{(C_{+}-\lambda_0)K_{+}}\ket{0}.
\end{equation}
The next step is to express the exponential term of $K_{+}\,(=a_0^{\dagger}\underline{a}_0^{\dagger})$ in terms of $b$-operators, which is done in (\ref{eq:ea+a+}). By replacing $\tau\to C_{+}-\lambda_0$, and $u\to u_0$, $v\to v_0$ there, we finally get 
\begin{equation}
\ket{\Psi(t)}\sim \sqrt{1-\abs{\lambda(t)}^2}\,e^{\lambda(t)b^{\dagger}\underline{b}^{\dagger}}\ket{0},
\label{eq:WF2}
\end{equation}
where we have put back the normalization factor explicitly, and 
\begin{equation}
\lambda(t)=\frac{C_{+}(t)-\lambda_{0}}{1-\lambda_{0}C_{+}(t)}=\frac{u_0C_{+}(t)-v_0}{u_0-v_0C_{+}(t)},
\label{eq:lambdat2a}
\end{equation}
with $\lambda_0=v_0/u_0$. 

Up to now we have obtained two equivalent forms for the same post-quench wavefunction: one is written in terms of $a_0$-operator with respect to its vacuum $\ket{\mathrm{GS}_0}$ [see (\ref{eq:WF1})], the other in terms of $b$-operator with respect to the vacuum $\ket{0}$ [see (\ref{eq:WF2})].

A few comments on $C_{+}(t)$ could be worth while. First we note that $C_{+}(t)$ in (\ref{eq:C+}) can also be written as
\begin{equation}
C_{+}(t)=\tilde{v}^*(t)/\tilde{u}^*(t),
\label{eq:C+2}
\end{equation}
where the definition of $\tilde{u}(t)$ and $\tilde{v}(t)$ can be found in (\ref{eq:uv-t}), with only difference of $\theta$ being replaced by $\tilde{\theta}$. In particular, when $g_0=0$, we have $u_0=1$, $v_0=0$, $\theta_0=0$, which leads to $\tilde{\theta}=\theta$, and $\tilde{u}(t)\to u(t)$, $\tilde{v}(t)\to v(t)$. Then we return to the familiar case of $0\to g$ quench, which has been addressed in \cite{Dora:2016} in detail. Actually (\ref{eq:WF1}) and (\ref{eq:C+2}) can be regarded as a direct application of the result in $0\to g$ quench to the general $g_0\to g$ quench, under the correspondence $b\to a_0$, $\ket{0}\to \ket{\mathrm{GS}_0}$, together with the transformation given by (\ref{eq:uvtilde}).

Secondly, we see that from (\ref{eq:C+}), the quantity $C_{+}(t)$ depends not only on $g$, but also on $g_0$ as well, since $\tilde{\theta}=\theta-\theta_0$. On the other hand, if $\lambda(t)$ in (\ref{eq:lambdat2a}) is concerned, the dependence of $g$ and $g_0$ there can be separated in a more explicit way. We can easily verify that
\begin{equation}
\lambda(t)=\frac{u_0v^*(t)-v_0u(t)}{u_0u^*(t)-v_0v(t)},
\label{eq:lambdat2b}
\end{equation}
where $u(t)$ and $v(t)$ are given by (\ref{eq:uv-t}). They appear as coefficients in the equation of motion for $b$-operator [see \ref{appC}], depending solely on the after-quench Hamiltonian $H(g)$. For $u_0$, $v_0$, they are determined by $g_0$ via (\ref{eq:uvtheta}).

From the above analysis, we see that starting from a pairing state with the form of (\ref{eq:GS2}), when the system is subject to a $g$-quench, the aftermath time-evolved wavefunction still adapts a simple pairing form. Depending whether the $a_0$-pairing or the $b$-pairing picture is chosen, the form of the wavefunction can be either (\ref{eq:WF1}) or (\ref{eq:WF2}). Indeed an observation can be simply made: for any given paired state that is the null state of a given class of Bogoliubov quasiparticle annihilation operators, it can be equally expressed as a paired state in terms of another class of Bogoliubov quasiparticle operators, with respect to their own corresponding vacuum state. Different classes of Bogoliubov operators are related via SU(1,1) (Bogoliubov) transformations as in (\ref{eq:uv}).



\section{Loschmidt echo and the entanglement in momentum space}\label{sec:LE}

One of the motivations of the present work for considering the general $\ g_0\to g$ quench is to examine an interesting observation made by D\'ora {\em et al} in \cite{Dora:2016}, which has been mentioned in the introduction part~\ref{sec:intro}. In that paper it has been shown that for the $0\to g$ quench about the Hamiltonian~(\ref{eq:H1}), the largest entanglement eigenvalue in the momentum space is always identical to the Loschmidt echo. The observation has been confirmed on various lattice models to which the Luttinger liquid theory can be applied, including the prestigious XXZ model in its gapless phase~\cite{Dora:2016}.

Here for general $g_0\to g$ quench, the Loschmidt echo (or the return probability) can be defined as usual as the overlap of the initial GS wavefunction and the final state wavefunction~\cite{Quan:2006,Dora:2013}, i.e., 
\begin{equation}
\mathcal{L}(t)=\abs{\braket{\Psi(0)}{\Psi(t)}}^2.
\label{eq:LE}
\end{equation}
From the expression of $\ket{\Psi(t)}$ in (\ref{eq:WF1}), we have
\begin{equation}
\mathcal{L}(t)=\abs{\braket{\mathrm{GS}_0}{\Psi(t)}}^2=1-\abs{C_+}^2
\label{eq:LE2}
\end{equation}
since $\ket{\Psi(0)}=\ket{\mathrm{GS}_0}$. Or more precisely, $\mathcal{L}(t)=\prod_{q>0}\left(1-\abs{C_{+}(t;q)}^2\right)$ if the $q$ dependence is restored. On the other hand, analogous to the discussion in \cite{Dora:2016}, it is straightforward to show that the largest eigenvalue of the reduced density matrix $\rho_{\mathrm{A}}(t)$ for the right-movers of $a_0$-Bogoliubov quasiparticles, denoted as $P_{\mathrm{max}}^{(1)}$ here, reads
\begin{equation}
P_{\mathrm{max}}^{(1)}=1-\abs{C_{+}(t)}^2=\abs{\tilde{u}(t)}^{-2},
\label{eq:Pmax1a}
\end{equation}
where $\tilde{u}(t)$ has been introduced in (\ref{eq:C+2}). It coincides with $\mathcal{L}(t)$ in (\ref{eq:LE2}), i.e.,
\begin{equation}
P_{\mathrm{max}}^{(1)}=\mathcal{L}(t)=\abs{\braket{\Psi(0)}{\Psi(t)}}^2.
\label{eq:Pmax1b}
\end{equation}
Thus similar to the $0\to g$ quench, the correspondence between the Loschmidt echo and the largest entanglement eigenvalue in the momentum space is established. 

However when it comes to a numerical simulation on a lattice model, it is the Fock space of the constituent particles that we are going to work with, whose vacuum is the empty state of these particles by definition. It remains open whether the low-lying entanglement spectrum obtained in this Fock space after tracing out, say, the $q<0$ partition is identical to the one obtained from the effective Hamiltonian~(\ref{eq:H1}). Indeed, from the results in the previous section, we note that the reduced density matrix $\rho_{\mathrm{A}}(t)$ and the subsequent entanglement spectrum can be evaluated either by the wavefunction~(\ref{eq:WF1}) or by (\ref{eq:WF2}), where the vacuum is the null state of either $a_0$ operators or $b$ operators.

We have addressed the situation with the wavefunction~(\ref{eq:WF1}) so far. For the other case, i.e., based on the wavefunction~(\ref{eq:WF2}), the entanglement between the left- and right-moving elemental excitations with respect to the $b$ operators can be evaluated as before. The largest entanglement eigenvalue reads 
\begin{equation}
P_{\mathrm{max}}^{(2)}=1-\abs{\lambda(t)}^2=\abs{u_0u(t)+v_0v(t)}^{-2},
\label{eq:Pmax2a}
\end{equation}
where the expressions of $u(t)$ and $v(t)$ (with respect to the $g$ value) can be found in (\ref{eq:uv-t}), and we note that here different from $P_{\mathrm{max}}^{(1)}$ in~(\ref{eq:Pmax1b}), 
\begin{equation}
P_{\mathrm{max}}^{(2)}=\abs{\braket{0}{\Psi(t)}}^2.
\label{eq:Pmax2b}
\end{equation}
The relation between $P_{\mathrm{max}}^{(1)}$ and $P_{\mathrm{max}}^{(2)}$ can be made transparent by employing Eq.~(\ref{eq:lambdat2a}) to give
\begin{equation}
P_{\mathrm{max}}^{(1)}=\frac{1-\lambda_0^2}{\abs{1+\lambda_0\lambda(t)}^2}\left(1-\abs{\lambda(t)}^2\right)=\frac{1-\lambda_0^2}{\abs{1+\lambda_0\lambda(t)}^2}P_{\mathrm{max}}^{(2)}.
\end{equation}

The above analysis shows that in general $g_0\to g$ quench, the correspondence between the Loschmidt echo defined in (\ref{eq:LE}) and the entanglement eigenvalue holds, only when the entanglement is measured of the left- and right-moving elementary excitations with respect to the initial GS of $H(g_0)$, i.e., $\ket{\mathrm{GS}_0}$. Here the excitations are nothing but the Bogoliubov $a_0$-quasiparticles. It is worthwhile asking the question of which entanglement indeed being evaluated in the numerical simulations on lattice models, and consequently what kind of wavefunction overlap the largest entanglement eigenvalue corresponds. We will address this issue in the next section by examining a spinless interacting fermionic lattice model, which is closely related to the prestigious XXZ model. 

\section{Numerical results}\label{sec:numerics}

The Hamiltonian of the $S=1/2$ XXZ model, which describes the Heisenberg-type interaction between neighboring $1/2$-spins along a chain, reads 
\begin{equation}
H=J\sum_j\left(S_{j+1}^xS_{j}^x+S_{j+1}^yS_{j}^y\right)+J_z\sum_jS_{j+1}^zS_{j}^z,
\label{eq:XXZ1}
\end{equation}
where $S_j^{\alpha}$s are spin operators at site $j$. Employing the Jordan-Wigner transformation that maps the spin operators into a set of fermionic ones with extra nonlocal string terms, the spin Hamiltonian can be recast into a system of spinless interacting fermions:
\begin{equation}
H=\frac{J}{2}\sum_j\left(c_{j+1}^{\dagger}c_{j}+\mathrm{H.c.}\right)
+J_z\sum_j\left(c_{j+1}^{\dagger}c_{j+1}-1/2\right)\left(c_j^{\dagger}c_j-1/2\right).
\label{eq:XXZ2}
\end{equation}
We see that the XX terms in the spin Hamiltonian are simply transformed into the hopping terms in the fermionic Hamiltonian. Meanwhile, the $J_z$-terms produce the scattering terms.

In the momentum space the Hamiltonian~(\ref{eq:XXZ2}) reads
\begin{equation}
H=\sum_k\left[J\cos(k)-J_z\right]c_k^{\dagger}c_k+\frac{J_z}{N}\sum_{k,p,q}\cos(q)c_{p-q}^{\dagger}c_pc_{k+q}^{\dagger}c_k, 
\label{eq:XXZ3}
\end{equation}
with a constant term $J_zN/4$ ($N$ is the lattice size) being dropped.
We may define a parameter $\Delta:=J_z/J$, which measures the strength of the uniaxial anisotropy along the $z$ direction. In the following discussions, we set $J = 1$. The energy and time are measured in units of $J$ and $J^{-1}$, respectively. It is well known that the XXZ model is critical (gapless) in the interval $\abs{\Delta}\le 1$, with a spin-singlet GS. Correspondingly, the GS of the Hamiltonian~(\ref{eq:XXZ3}) in this regime takes place at the half filling, and its low-energy physics can be well described by the Luttinger liquid, e.g., see \cite{Voit:1995ig, Giamarchi:2004, Franchini:2017}.

The way from the fermionic Hamiltonian~(\ref{eq:XXZ3}) (abbreviated as {$H_{\mathrm{F}}$}) to the paired bosonic Hamiltonian~(\ref{eq:H1}) (abbreviated as {$H_{\mathrm{B}}$}) can be established via the bosonization technique~\cite{Gogolin:2004vy}. Without going into technical details, we would like to emphasize that here the umklapp term is dropped since it is irrelevant in the gapless phase, and the $g_2$ term, which describes the forward scattering processes on both sides of the Fermi surface, is kept explicitly. In the Hamiltonian~(\ref{eq:H1}) it appears as the  pairings of the bosonic $b$ operators with opposite momenta~\cite{Giamarchi:2004}. The constant quantity $g$ in (\ref{eq:wg}) is indeed identical to $g_2$. The Luttinger parameter reads $K=\sqrt{(v-g)/(v+g)}$ in {$H_{\mathrm{B}}$}~\cite{Dora:2016}, the precise value of which in the XXZ model is given by $K=\pi/2(\pi-\arccos\Delta)$ via the Bethe ansatz. $K\in[1/2,+\infty)$ in the gapless regime $\abs{\Delta}\le 1$, and the noninteracting point $K=1$ corresponds to the XY limit ($\Delta=0$).

\begin{figure}
\centering
\includegraphics[width=0.8\textwidth]{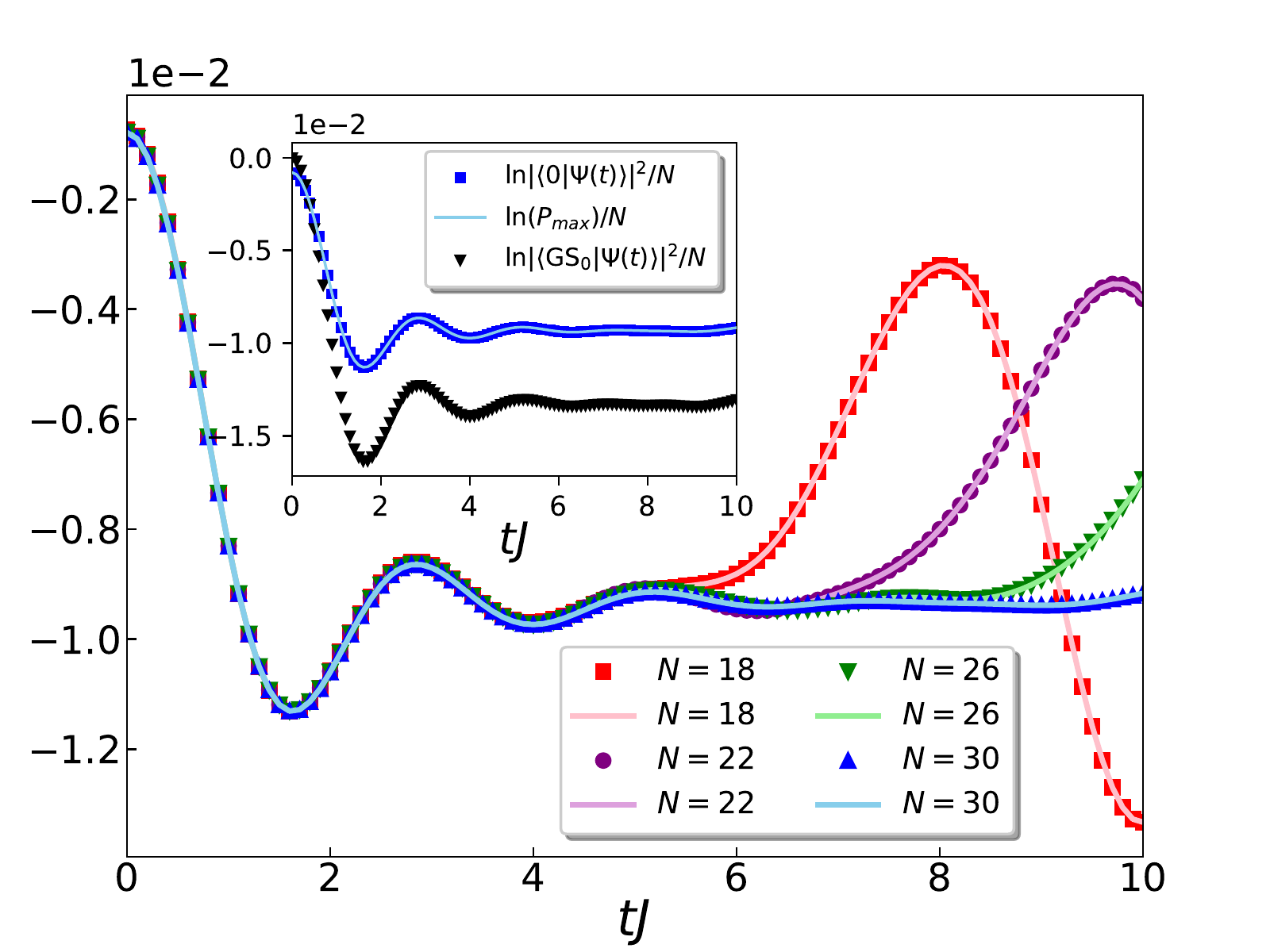}
\caption{(Color online) The succeeding time evolution of the logarithms of $\abs{\braket{0}{\Psi(t)}}^2$ (symbols) and the largest entanglement eigenvalue $P_{\mathrm{max}}(t)$ (solid curves) (divided by the lattice size $N$) after the quench of $\Delta_0=-0.2\to\Delta=0.4$ at $t=0$ over the Hamiltonian~(\ref{eq:XXZ3}). Here $\ket{0}$ is the noninteracting GS at $\Delta=0$, and $\Psi(t)$ is the time-evolved wavefunction. $P_{\mathrm{max}}(t)$ is determined by the reduced density matrix $\rho_{\mathrm{A}}(t)$ after tracing out the left-moving part in $\Psi(t)$. The results for $N=18,22,26,30$ are presented. The inset shows the comparison between $P_{\mathrm{max}}(t)$, and two kinds of wavefunction overlap for $N=30$, i.e., $\abs{\braket{0}{\Psi(t)}}^2$ and $\abs{\braket{\mathrm{GS}_0}{\Psi(t)}}^2$, respectively. As mentioned in the main text, $\ket{\mathrm{GS}_0}\,(=\ket{\Psi(0)})$ is the GS of the system at $\Delta_0=-0.2$. The deviation of $\abs{\braket{\mathrm{GS}_0}{\Psi(t)}}^2$ from the other two is clearly recognized.}
\label{fig1}
\end{figure}

In the numerical simulation, we work directly on the interacting fermionic Hamiltonian~(\ref{eq:XXZ3}) at half filling in momentum space by the exact diagonalization (ED). The time-evolved wavefunction $\ket{\Psi(t)}$ is obtained by the standard time-dependent Lanczos method~\cite{Prelovsekbook}. The reduced density matrix $\rho_{\mathrm{A}}(t)$ with respect to $\ket{\Psi(t)}$ and the consequent entanglement spectrum are procured accordingly, after tracing out $q<0$ part in the momentum space. The results for the quench of $(\Delta_0=-0.2)\to(\Delta=0.4)$ on various lattice sizes ($N=18,22,26,30$) are presented in figure \ref{fig1}. In the main figure, the logarithms of both $\abs{\braket{0}{\Psi(t)}}^2$ and the largest entanglement eigenvalue $P_{\mathrm{max}}(t)$ are shown. The inset displays a comparison with that of $\abs{\braket{\mathrm{GS}_0}{\Psi(t)}}^2$ for $N=30$. Here following the previous conventions, $\ket{0}$ is the noninteracting GS (at $\Delta=0$), and $\ket{\mathrm{GS}_0}$ is the GS at $\Delta_0$. The quench takes place at $t=0$, with $\ket{\Psi(0)}=\ket{\mathrm{GS}_0}$. For a given $N$ the agreement between $P_{\mathrm{max}}(t)$ and $\abs{\braket{0}{\Psi(t)}}^2$ holds well during the time evolution. On the other hand, the deviation from $\abs{\braket{\mathrm{GS}_0}{\Psi(t)}}^2$ is evident, as shown in the inset of figure \ref{fig1}. It indicates that it is $P_{\mathrm{max}}^{(2)}$ in {$H_{\mathrm{B}}$}, rather than $P_{\mathrm{max}}^{(1)}$ is compatible with $P_{\mathrm{max}}$ we have calculated in {$H_{\mathrm{F}}$}. Remind that $P^{(2)}$ is about the momentum-space entanglement in terms of $b$ operators. Additionally, we observe that following a rapid decrease in the early stage, the later-time oscillation of these time-dependent quantities, which is due to the discreteness of the energy spectrum in finite systems, is evidently suppressed with increasing lattice size. We note that the short transient time before the saturation taking place is around $4\,J^{-1}$ from the numerical results, consistent with the estimation from the effective field theory~\cite{Dora:2016}. 

The above result can be understood as a natural ramification of the {$H_{\mathrm{F}}\leftrightarrow H_{\mathrm{B}}$} correspondence. Let us first consider the equilibrium case and take a look at the simple noninteracting case. It means $\Delta(:=J_z/J)=0$ in {$H_{\mathrm{F}}$}, or equivalently, $g=0$ in {$H_{\mathrm{B}}$} ($K=1$). In either situation, we have free particles with momenta as good quantum numbers. Both ground states are disentangled in momentum space, with only a single nonzero entanglement eigenvalue $P=1$. As the interaction is turned on, which means $\Delta\neq 0$ on the fermionic side, we can expect that the entanglement between the $q>0$ and $q<0$ partitions will develop due to the interaction, yet it may not be easy to write down the wavefunction explicitly to figure it out. On the other hand, compared to the rather complicated situation on the fermionic side, the momentum-space entanglement can be much more easily estimated from the GS wavefunction of the bosonic counterpart [see (\ref{eq:GS})], where the entanglement between $(q,-q)$-pairs with respect to the $b$-operators can be readily read out. 

If we agree with the effectiveness of the Luttinger liquid description of {$H_{\mathrm{F}}$}~\cite{Karrasch:2012, Collura:2015}, we can expect that in the nonequilibrium case, the time evolution of the entanglement in the general $\Delta$-quench over {$H_{\mathrm{F}}$} should follow the features of its bosonic partner, as long as the $\Delta\leftrightarrow g$ relation is sustained. Thus, the arbitrariness in the entanglement calculation for the $g_0\to g$ quench in {$H_{\mathrm{B}}$} can be removed: we need to stick to the $b$-representation consistently and adopt the time-evolved wavefunction in the form of (\ref{eq:WF2}), in order to make reasonable a comparative study with its fermionic counterpart. Specifically in this section, we have ratified the speculation on the relation between the time-dependent largest entanglement eigenvalue and the overlap of wavefunctions for general quenches, which is drawn from the analysis on the bosonic case [see (\ref{eq:Pmax2b})], for the interacting fermionic lattice model.

\section{Discussion and conclusion}\label{sec:conclusion}

In this work, we first consider a paired bosonic model, which serves as an effective Hamiltonian for spinless Luttinger liquids. By employing the $\mathsf{su}(1,1)$ algebra, we derive the exact form of the time-evolved wavefunction under general interaction quenches on pairing strength. The method we employ and the results are general as long as the pairing takes place in the individual $(q,-q)$ channels. The application to the fermionic case is straightforward, with only the $\mathsf{su}(1,1)$ algebra being replaced by $\mathsf{su}(2)$.

Based on the analytic form of the wavefunction, we re-examine under the general quench setup, the correspondence between the entanglement spectrum and the wavefunction overlap, including the Loschmidt echo. We note that by employing different sets of Bogoliubov quasiparticles, the expression for a given time-evolved post-quench quantum state can be different. It leads to an ambiguity in measuring the entanglement of paired bosons. However, by working numerically on an interacting spinless fermionic model that is closely related to the XXZ spin chain, we find that the ambiguity can be removed if the concurrence between the fermionic and bosonic descriptions is required. More specifically, we confirm that on the fermionic lattice model, the time-dependent largest entanglement eigenvalue in the momentum space coincides with the overlap between the time-evolved wavefunction and its noninteracting GS, rather than the initial state when the quench starts. 

It is well known that quantum entanglement plays a key role in understanding the nature of quantum world, especially of many-body systems~\cite{Amico:2008en, Laflorencie:2016}. The entanglement entropy and entanglement spectrum have been routinely investigated in the study. Up to the present most studies predominantly focus on real or orbital space entanglement. For many gapped systems, the congruity between the energy spectrum of the edge states and the entanglement spectrum obtained in this manner has been established~\cite{Chandran:2011, QiXL:2012, Senthil:2012}. Nevertheless for critical (gapless) systems, due to the absence of energy gap, the intrinsic entanglement might not be evidenced by cuttings in real space, which is corroborated by the observation of the slow decay of eigenvalue spectrum of the reduced density matrix in the real-space density-matrix renormalization group (DMRG)~\cite{Schollwock:2005}. 

Consequently, the issue of the entanglement structure in momentum space in gapless systems was raised and explored, first for spin chains~\cite{Thomale:2010, Lundgren:2014, Lundgren:2016, Berganza:2016}, then for other correlated systems, including, e.g., the Hubbard model~\cite{Ehlers:2015} and the disordered fermionic systems~\cite{Hughes:2013, Lundgren:2019}. In the framework of the Luttinger liquid theory, where the elementary excitations fall into left- and right-moving (weakly) interacting particles, it is particularly natural to consider partitioning the many-body Hilbert space by momentum. It is quite significant that the correspondence between the largest momentum-space entanglement eigenvalue and the wavefunction overlap in nonequilibrium dynamics, which is rather plain in the effective bosonic model, has been verified by the numerical observations on interacting lattice models. It suggests that the entanglement in momentum space can furnish a new perspective for the study of nonequilibrium dynamics of quantum critical many-body systems. 

Furthermore, from the perspective of the development of numerical algorithms, the investigation on the momentum-space entanglement can help to improve the performance and identify the most fruit-bearing applications of the (time-dependent) momentum-space DMRG~\cite{Xiang:1996ex, Schollwock:2005}. For example, there have been reports on the persistence of the entanglement gap in momentum space beyond the physically critical points, and the saturation of the entanglement entropy in nonequilibrium dynamics~\cite{Lundgren:2016, Dora:2016, Lundgren:2019}. The DMRG working in the momentum space might have potential to outperform its traditional real-space counterpart in these regimes. 

Before concluding this section, we would like to mention that as far as the correspondence in quench dynamics is concerned, there may be some interesting issues remained to be investigated, which are beyond the scope of the present work. For instance,  we notice in the numerical simulations that the correspondence we have discussed largely holds even beyond the regime of the gapless phase where the Luttinger liquid theory works well. To be precise, for the quenches crossing through the Berezinsky-Kosterlitz-Thouless (BKT) critical point ($\Delta=1$), which separates the gapless phase from the Ising antiferromagnetic gapped phase, the correspondence sustains up to some critical value of $\Delta>1$ (e.g., around $\Delta\sim 1.6$ for $\Delta_0=0.5$). In contrast, for the quenches moving in the opposite direction, i.e., from the critical phase to the gapped ferromagnetic phase cross the first-order-like critical point ($\Delta=-1$), the behaviors are quite different, and the agreement quickly dissolves. To study and understand the extent of validity of the predictions from the Luttinger liquid theory can be interesting.

\section*{Acknowledgments}
H.L. acknowledge support from the National Natural Science Foundation of China (Grants No. 11474136, No. 11874187, and No. 12047501).

\appendix
\numberwithin{equation}{section}
\section{The Baker-Campbell-Hausdorff relation}\label{appA}

In this appendix, we show that the operator $U(\tau)$ with the form as
\begin{equation}
U(\tau)=\exp\left[\tau\left(xK_{+}+yK_{-}+zK_{0}\right)\right]:=\exp(\tau H),
\label{eq:U}
\end{equation}
can be expressed as 
\begin{equation}
U(\tau)=e^{C_{+}(\tau)K_{+}}e^{C_{0}(\tau)K_{0}}e^{C_{-}(\tau)K_{-}},
\label{eq:U2}
\end{equation}
where $K_{0}$ and $K_{\pm}$ satisfy the $\mathsf{su}(1,1)$ Lie algebra [see (\ref{eq:su11})]~\cite{Gilmore:1974}, and $x$, $y$, $z$ are constants. The coefficients in the exponents in (\ref{eq:U2}), which satisfy the initial condition of $C_{+}(0)=C_{-}(0)=C_{0}(0)=0$, are given by
\begin{eqnarray}
C_{-}(\tau)=\frac{e^{\tau yr_2}-e^{\tau yr_1}}{r_2e^{\tau yr_2}-r_1e^{\tau yr_1}},\qquad C_{+}(\tau)=r_1r_2C_{-}, \nonumber \\
C_{0}(\tau)=\tau y\left(r_1+r_2\right)-2\ln\frac{r_2e^{\tau yr_2}-r_1e^{\tau yr_1}}{r_2-r_1},
\label{eq:Cs}
\end{eqnarray}
where $r_{1,2}$ are the roots of a quadratic equation of $t$: $yt^2+zt+x=0$. Note that the solution, though quite general, is not applicable in all situations, e.g., it fails when $r_1=r_2$. We will discuss a specific example for this case in~\ref{appB}.

Following the procedure detailed in \cite{Truax:1985}, the derivation of the above result is quite straightforward. First note that the operator $U(\tau)$ satisfies a differential equation as
\begin{equation}
\left[\partial\tau U(\tau)\right]U^{-1}(\tau)=xK_{+}+yK_{-}+zK_{0}=H.
\end{equation}
Substituting the second form of $U(\tau)$~(\ref{eq:U2}) to the equation, and using the commutation relations of $K$s, we can obtain the differential equations for $C$-coefficients as
\begin{equation}
\left\{\begin{array}{l}\dot{C}_{+}-\dot{C}_0C_{+}+\dot{C}_{-}C^2_{+}e^{-C_0}=x, \\
\dot{C}_{-}e^{-C_{0}}=y,\\
\dot{C}_{0}-2\dot{C}_{-}C_{+}e^{-C_{0}}=z. 
\end{array}
\right.
\label{eq:diffCs}
\end{equation}
The above equations, although nonlinear, can be integrated analytically. For instance, the equation for $C_{+}$ can be written as
\begin{equation}
\dot{C}_{+}-yC_{+}^2-zC_{+}=x,
\label{eq:C+3}
\end{equation}
which can be solved by separation of variables: $\ud C_{+}/(x+yC_{+}^2+zC_{+})=\ud \tau$. In the case of the quadratic equation $yt^2+zt+x=0$ having distinct roots, we get the solution in~(\ref{eq:Cs}).

\section{The decomposition of $\exp({\tau a^{\dagger}\underline{a}^{\dagger}})$}\label{appB}

In this section as a supplement to the discussion in~\ref{appA}, let us consider an exponential form of the pair creation operators of the Bogoliubov quasiparticles, $U(\tau):=\exp({\tau a^{\dagger}\underline{a}^{\dagger}})$. Note that $a^{\dagger}=ub^{\dagger}+v\underline{b}$ [see (\ref{eq:uv}), and we also confine ourselves to real $u,\,v$], we have
\begin{equation}
U(\tau)=e^{\tau(ub^{\dagger}+v\underline{b})(u\underline{b}^{\dagger}+vb)}:=e^{\tau H},
\end{equation}
with
\begin{equation}
H=u^2K_{+}+v^2K_{-}+2uvK_{0}.
\end{equation}
The $K$ operators, as a realization of the $\mathsf{su}(1,1)$ Lie algebra, are given by
\begin{equation}
K_0=\frac{1}{2}(b^{\dagger}b+\underline{b}\underline{b}^{\dagger}), \qquad 
K_{+}=b^{\dagger}\underline{b}^{\dagger},\qquad K_{-}=b\underline{b}.
\label{eq:Kvsb}
\end{equation}
Though we cannot apply directly the result in (\ref{eq:Cs}) since here $r_{1}=r_{2}=-u/v$, the coefficients in~(\ref{eq:U2}) can still be solved as following. From (\ref{eq:C+3}), we have
\begin{equation*}
\dot{C}_{+}=v^2C_{+}^2+2uvC_{+}+u^2=(vC_{+}+u)^2.
\end{equation*}
Then 
\begin{equation}
C_{+}(\tau)=\frac{u^2\tau}{1-uv\tau},\qquad C_{-}=\frac{C_{+}}{r_1r_2}=\frac{v^2\tau}{1-uv\tau},
\end{equation}
and
\begin{equation}
\quad C_{0}=-\ln(y/\dot{C}_{-})=-2\ln(1-uv\tau).
\end{equation}
We obtain consequently
\begin{equation}
\exp({\tau a^{\dagger}\underline{a}^{\dagger}})
=e^{\frac{u^2\tau}{1-uv\tau}K_{+}}e^{-2\ln(1-uv\tau)K_{0}}e^{\frac{v^2\tau}{1-uv\tau}K_{-}}.
\label{eq:ea+a+}
\end{equation}

As an application of (\ref{eq:ea+a+}), let us consider the GS of the Hamiltonian~(\ref{eq:H1}), which, in our simplified notation, reads
\begin{equation}
\ket{\mathrm{GS}}=\frac{1}{u}e^{-\frac{v}{u}b^{\dagger}\underline{b}^{\dagger}}\ket{0}.
\label{eq:GS2b}
\end{equation}
We say that the GS can also be expressed as 
\begin{equation}
\ket{\mathrm{GS}}=ue^{-\frac{v}{u}a^{\dagger}\underline{a}^{\dagger}}\ket{0}.
\label{eq:GS3}
\end{equation}
The conclusion can be verified in a straight way by means of replacing $\tau$ in (\ref{eq:ea+a+}) with $-v/u$, and using the fact that $\ket{0}$ is the vacuum state of $b$, i.e., $b\ket{0}=0$. 

At the end of the section, an extra comment might be interesting. Note that (\ref{eq:GS3}) can be rewritten as
\begin{equation}
\ket{0}=\frac{1}{u}e^{\frac{v}{u}a^{\dagger}\underline{a}^{\dagger}}\ket{\mathrm{GS}}.
\label{eq:GS3b}
\end{equation}
Here we see that there is some kind of duality between (\ref{eq:GS2b}) and (\ref{eq:GS3b}). For the vacuum of the constituent particles, i.e., $\ket{0}$, we can equally regard it as a kind of pairing state of Bogoliubov quasiparticles as well, with respect to its own vacuum state.

\section{The equation of motion for $b$-operators}\label{appC}

Consider the operator $b$ in the Heisenberg picture $b(t)=e^{iHt}\,b\,e^{-iHt}$, with $H$ given by (\ref{eq:H1}). Then
\begin{equation}
b(t)=e^{iHt}(ua-va^{\dagger})e^{-iHt}
=ue^{-i\epsilon t}a-ve^{i\epsilon t}a^{\dagger}:=u(t)b+v^*(t)b^{\dagger},
\label{eq:bt}
\end{equation}
where 
\begin{eqnarray}
u(t)&=&u^2e^{-i\epsilon t}-v^2e^{i\epsilon t}=\cos(\epsilon t)-i\sin(\epsilon t)\cosh\theta,\nonumber \\
v(t)&=&uv\left(e^{i\epsilon t}-e^{-i\epsilon t}\right)=i\sin(\epsilon t)\sinh\theta.
\label{eq:uv-t}
\end{eqnarray}
In the above derivation, the Hamiltonian in another form of (\ref{eq:H3}) and the Bogoliubov transformation (\ref{eq:uv}) (and (\ref{eq:uvtheta})) have been used. 

We note that with $u(t)$, $v(t)$ defined above, the time-evolved wavefunction for the $0\to g$ quench can be written as $\Psi(t)\sim \frac{1}{\abs{u^*(t)}}e^{\frac{v^*(t)}{u^*(t)}b^{\dagger}\underline{b}^{\dagger}}\ket{0}$~\cite{Dora:2016}. It suggests that in order to obtain $\ket{\Psi(t)}$, there could be some convenient way by employing the time-dependent features of $b(t)$. This is true actually. In the following we show that by assuming $\ket{\Psi(t)}=\gamma(t)e^{\lambda(t)b^{\dagger}\underline{b}^{\dagger}}\ket{0}$, we can quickly figure out the coefficient $\lambda(t)$. First from (\ref{eq:bt}), we have
\begin{equation*}
b(t)\ket{0}=v^*(t)b^{\dagger}\ket{0};
\end{equation*}
On the other hand, from the definition of $b(t)$, we have
\begin{eqnarray*}
b(t)\ket{0}&=&e^{iHt}be^{-iHt}\ket{0}=e^{iHt}b\left[\gamma(t)e^{\lambda(t)b^{\dagger}\underline{b}^{\dagger}}\ket{0}\right] 
=\gamma e^{iHt}e^{\lambda b^{\dagger}\underline{b}^{\dagger}}(b+\lambda b^{\dagger})\ket{0} \\
&=&e^{iHt}(\lambda b^{\dagger})\gamma e^{\lambda b^{\dagger}\underline{b}^{\dagger}}\ket{0}
=e^{iHt}(\lambda b^{\dagger})e^{-iHt}\ket{0}=\lambda b^{\dagger}(t)\ket{0}=\lambda u^*(t)b^{\dagger}\ket{0}.
\end{eqnarray*}
The comparison of these two equations produces 
\begin{equation}
\lambda(t)=v^*(t)/u^*(t).
\label{eq:lambdat1}
\end{equation}
Actually by employing the same strategy with the same assumption about $\ket{\Psi(t)}$, we can show that for the general $g_0\to g$ quench, the coefficient $\lambda(t)$ in the exponent is
\begin{equation}
\lambda(t)=\frac{v^*(t)-\lambda_0u(t)}{u^*(t)-\lambda_0v(t)},
\label{eq:lambdat2c}
\end{equation}
where $\lambda_0:=v_0/u_0$, as appeared in (\ref{eq:GS2}). It is identical to the result~(\ref{eq:lambdat2b}) in the main text, which has been derived in a somehow intricate but rigorous way by employing the $\mathsf{su}(1,1)$ algebra. 

\vspace*{2mm}


\end{document}